\documentclass[a4paper,fleqn,usenatbib]{mnras}
\usepackage{graphicx}	% Including figure files
\usepackage{amsmath}	% Advanced maths commands
\usepackage{amssymb}	% Extra maths symbols
\usepackage{multicol}        % Multi-column entries in tables
\usepackage{bm}		% Bold maths symbols, including upright Greek
\usepackage{pdflscape}	% Landscape pages
\usepackage{xcolor}

\usepackage[T1]{fontenc}
\usepackage{ae,aecompl}

\usepackage{newtxtext,newtxmath}
\title[Polarized Dust Emission in Arp 220]{Polarized Dust Emission in Arp220: Magnetic Fields in the Core of an Ultraluminous Infrared Galaxy}
\author[D.L. Clements et al.]{
D.L. Clements$^{1}$\thanks{Contact e-mail: d.clements@imperial.ac.uk},
Qizhou Zhang$^{2}$, K. Pattle$^3$,
G. Petitpas$^{2}$, Y. Ding$^{1}$, J. Cairns$^{1}$\newauthor
\\
$^{1}$Imperial College London, Prince Consort Road, London SW7 2AZ, UK\\
$^{2}$Center for Astrophysics $|$ Harvard \& Smithsonian, 60 Garden Street, Cambridge, MA 02138 \\
$^{3}$Department of Physics and Astronomy, University College London, Gower Street, London WC1E 6BT, UK\\
}\date{}
\pagerange{\pageref{firstpage}--\pageref{lastpage}}
\pubyear{}

\begin{document}

\maketitle

\label{firstpage}

\begin{abstract}
Arp 220 is the prototypical Ultraluminous Infrared Galaxy (ULIRG), and one of the brightest objects in the extragalactic far-infrared sky. It is the result of a merger between two gas rich spiral galaxies which has triggered starbursting activity in the merger nuclear regions. Observations with the Submillimeter Array centred at a frequency of 345 GHz and with a synthesised beamsize of $0.77\times0.45$ arcseconds were used to search for polarized dust emission from the nuclear regions of Arp 220. Polarized dust emission was clearly detected at 6\,$\sigma$ significance associated with the brighter, western nucleus, with a peak polarization fraction of 2.7$\pm 0.35$ per cent somewhat offset from the western nucleus.  A suggestive 2.6\,$\sigma$ signal is seen from the fainter eastern nucleus. The dust emission polarization is oriented roughly perpendicular to the molecular disk in the western nucleus suggesting that the magnetic field responsible is orientated broadly in the plane of the disk, but may be being reordered by the interaction between the two nuclei.  Unlike more evolved interacting systems, we see no indication that the magnetic field is being reordered by the outflow from the western nucleus. These observations are the first detection of dust polarization, and thus of magnetic fields, in the core of a ULIRG.
\end{abstract}

\begin{keywords}
galaxies: magnetic fields; galaxies: interactions; galaxies: starburst; submillimetre: galaxies; galaxies: individual: Arp220
\end{keywords}

\section{Introduction}

Magnetic fields are a key component of the interstellar medium (ISM) of galaxies and are involved in numerous processes including star formation, mass loss and the jets associated with active galactic nuclei (AGN) (see eg. \citet{b13} and references therein). Dust grains, another key component of the ISM, will align themselves with magnetic fields when aspherical, leading to polarized thermal emission in the far-IR and submm \citep{b13}.  Observations of polarized dust emission at these long wavelengths can thus provide insights into the magnetic fields in galaxies \citep{l03}. Observations of galactic magnetic fields can be used to test dynamo theory, which predicts the timescales for the coherent ordering of magnetic fields in galaxies. Studies of the flow patterns of diffuse ionised gas suggest that magnetic fields can affect the evolution of galaxy morphology (e.g. spiral arm formation), while major mergers are thought to enhance turbulent magnetic fields \citep{b13}. This means that observations of galactic magnetic fields can provide insights into both the merger history and morphological evolution of galaxies. In particular, \citet{w21} suggest that, while the presence of magnetic fields does not appear to affect global merger remnant properties, galactic magnetic fields can produce significant differences in the structural properties of merger remnants, with magneto-hydrodynamic (MHD) simulations producing large disc galaxies with spiral structures, compared to the more compact merger remnants predicted by hydrodynamic simulations. Magnetic fields may also have an impact on feedback processes, potentially leading to increased nuclear star formation rates and supermassive black hole (SMBH) accretion rates in the centres of merging galaxies \cite{w23}. The investigation of magnetic fields in merging galaxies is therefore important not only for understanding the processes underway in individual objects, but also for understanding the overall evolutionary history of galaxies and SMBHs.

Despite the potential impact of magnetic fields on galaxy structure, submm observations of polarization in extragalactic sources remain sparse. \citet{g02} published the first galaxy-averaged detection of submm polarization, finding a polarization fraction of $\sim0.4$ per cent at 850$\mu$m in the nearby starburst galaxy M82. This has since been followed up by higher resolution observations in the far-IR from SOFIA (Stratospheric Observatory for Infrared Astronomy, \citealt{j19}) and deeper observations in the submm \citep{p21} which have shown that M82 has one magnetic field component that aligns with the disk of the galaxy, and another aligned perpendicular to the disk associated with hotter dust and thought to come from material entrained in a massive polar outflow. Limited samples of nearby galaxies have since been detected in polarized dust emission by SOFIA (eg. \citealt{b23}, \citealt{l20}, and references therein), but the sample sizes remain small. At higher redshifts individual extreme sources have been studied and detected in polarized dust emission at $z=2.6$ \citep{g23}, and $z=5.6$ \citep{c24} with polarization fractions comparable to what has been seen locally. Statistical studies of dust polarization in a large number of galaxies using Planck data \citep{b17} suggest average fractional polarizations of $\sim3$ per cent for the population, and raise the interesting prospect that the integrated polarized dust emission of cosmic infrared background (CIB) galaxies might be a significant, if not dominant, foreground for future cosmic microwave background searches for B-mode polarzsation \citep{l20}. 

\begin{figure}
\includegraphics[width=8cm]{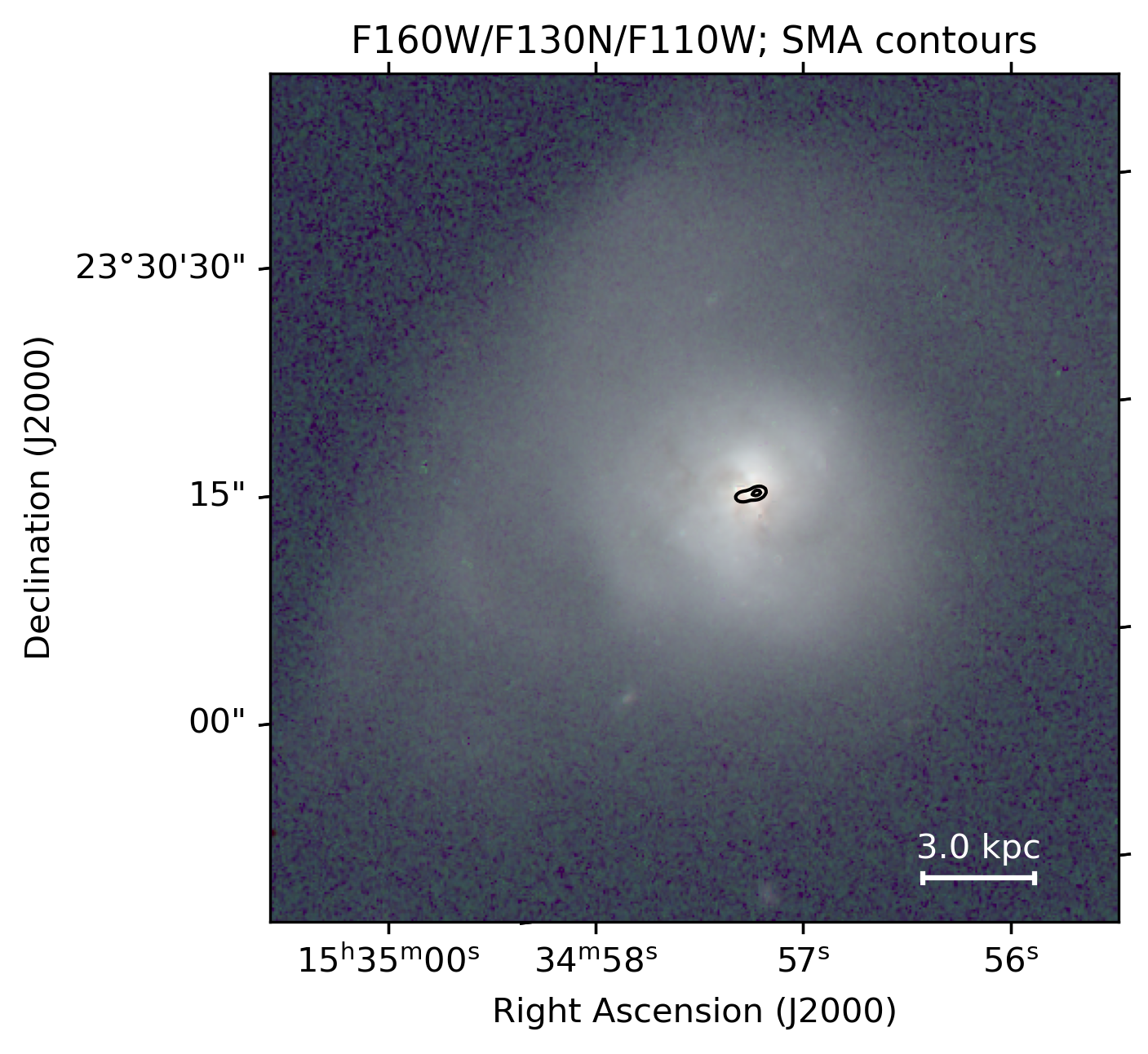}
\caption{Three colour near-IR image of Arp220 from the Hubble Space Telescope \citep{p20} showing the large scale structure of the merging system. White contours indicate the SMA continuum 340 GHz image obtained by our observations. Background spikes found in the archival data are suppressed by sigma-clipping at $3\sigma$ in $7 \times 7$ pixel boxes across the field using the astropy sigma\_clip routine.  A maximum of 10 iterations of sigma-clipping were performed in each box.}
\label{fig:opt}
\end{figure}

The main issues making further studies of dust polarization in galaxies difficult are sensitivity and angular resolution, as well as the demise of SOFIA, which had become the leading observatory for far-IR dust polarization studies. The sensitivity aspect of this problem suggests that the brightest far-IR/submm sources in the local universe would potentially make the best targets for further extragalactic dust polarization observations. Among these are the local Ultraluminous Infrared Galaxies (ULIRGs, see eg. \citealt{s96} and references therein), the brightest of which is the well known merging galaxy Arp 220, which is the target of the observations discussed here. Arp 220 is the merger of two gas rich spiral galaxies (see Figure \ref{fig:opt} \citealt{r11}), and hosts a massive starburst forming stars at a rate of $\sim100$ M$_{\odot}$/yr (eg. \citealt{d98}). This star formation is concentrated into two nuclei at its centre separated roughly east-west by $\sim1$ arcsecond corresponding to a physical distance of 300 pc at its redshift of $z=0.018$ (see eg. \citealt{s09, s97}). Each nucleus is surrounded by a rotating gas disk of $\sim 100$pc in extent, with the rotation axes of these disks misaligned \citep{s99}. The combined flux density of the two nuclei at 345 GHz is 680$\pm$100mJy, The western nucleus is about 3 times brighter than the eastern nucleus and is thought to contain an AGN \citep{c02, d07}.

Previous observations in search of dust polarization in Arp220 have set upper limits of 1.54 per cent \citep{s07}. However, these observations were taken with the Submillimetre Common User Bolometer Array (SCUBA,  \citealt{h99}) and are thus averaged over a 15 arcsecond beam. If the two nuclei have different polarizations then this would dilute the polarization, potentially to the level of non-detection. We here present the results of submm polarization observations of Arp220 at subarcsecond resolution using the Submilliemeter Array (SMA, \citealt{h04}). These are capable of resolving the separate nuclei and thus avoiding this dilution problem. If a significant polarization fraction can be found in the nuclei of Arp220, this would suggest that the magnetic field strengths can be enhanced in the inner regions of merging galaxies as suggested by \citet{w21}. Alternatively, if the polarization fraction remains undetectable even in the individual nuclei of Arp220, this suggests that the magnetic fields in merging nuclei may be much more complex and disordered than predicted, providing interesting constraints on the ordering of galactic magnetic fields.

The rest of this paper is structured as follows: in the next section we describe the SMA observations and data reduction. In the subsequent section we present the results of our observations. This is followed by discussion of these results, and then by our conclusions. Throughout this paper we assume a standard cosmology with $H_0$ = 67.74 km s$^{-1}$ Mpc$^{-1}$, $\Omega_{\Lambda}$ = 0.69 and $\Omega_m$ = 0.31.

\section{Observations and Data Reduction}

Observations of Arp 220 were carried out on 2022 May 25 using the Submillimeter Array (SMA). Six of the eight antennae were used in SMA's extended configuration during the observations, the other antennae being unavailable, providing projected baselines from 22.5 to 226 m. With an IF frequency of 4 to 16 GHz, the spectral coverage was from 331 to 343 GHz for the lower sideband (LSB) and 351 to 363 GHz for the upper sideband (USB) at a uniform resolution of 140 kHz across the entire spectral coverage. The weather conditions were excellent, with a precipitable water vapour (PWV) of 1 mm, and the zenith atmospheric opacity of 0.05 at 225 GHz.

Both 345 GHz (RxA) and 400 GHz (RxB) receivers were used during the observations. The quarter waveplates equipped in each antenna convert the incoming linearly polarized signals to circular ones \citep{m08}. Therefore, the two receivers  sample the left-hand and right-hand circular polarization L and R, respectively. The cross correlation produces the four polarization products, i.e., RR, LL, RL and LR, for each antenna pair. 3C345 and Bl-Lac were used  as time-dependent gain, and bandpass calibrators, respectively. Neptune and Callisto were observed for flux calibrations. The instrumental leakage calibration was performed using 3C345. Visibility data was first calibrated for flux, gain and bandpass using MIR/IDL subset, and then exported to MIRIAD \citep{s95} for polarization leakage calibration, and imaging. The continuum data for Arp 220 were constructed by vector averaging the spectral data in the LSB and USB, respectively. The synthesized beam of the combined continuum image is 0".77 x 0".45 with a position angle of -90$^{\circ}$. The one s$\sigma$ rms noise is 3 mJy per beam in the Stoke I image, and 0.5 mJy in the Stokes Q/U images. The pointing center for Arp 220 is 15:34:57.26,  $+$23:30:11.30 (J2000).

\begin{figure*}[h]
\includegraphics[width=18cm]{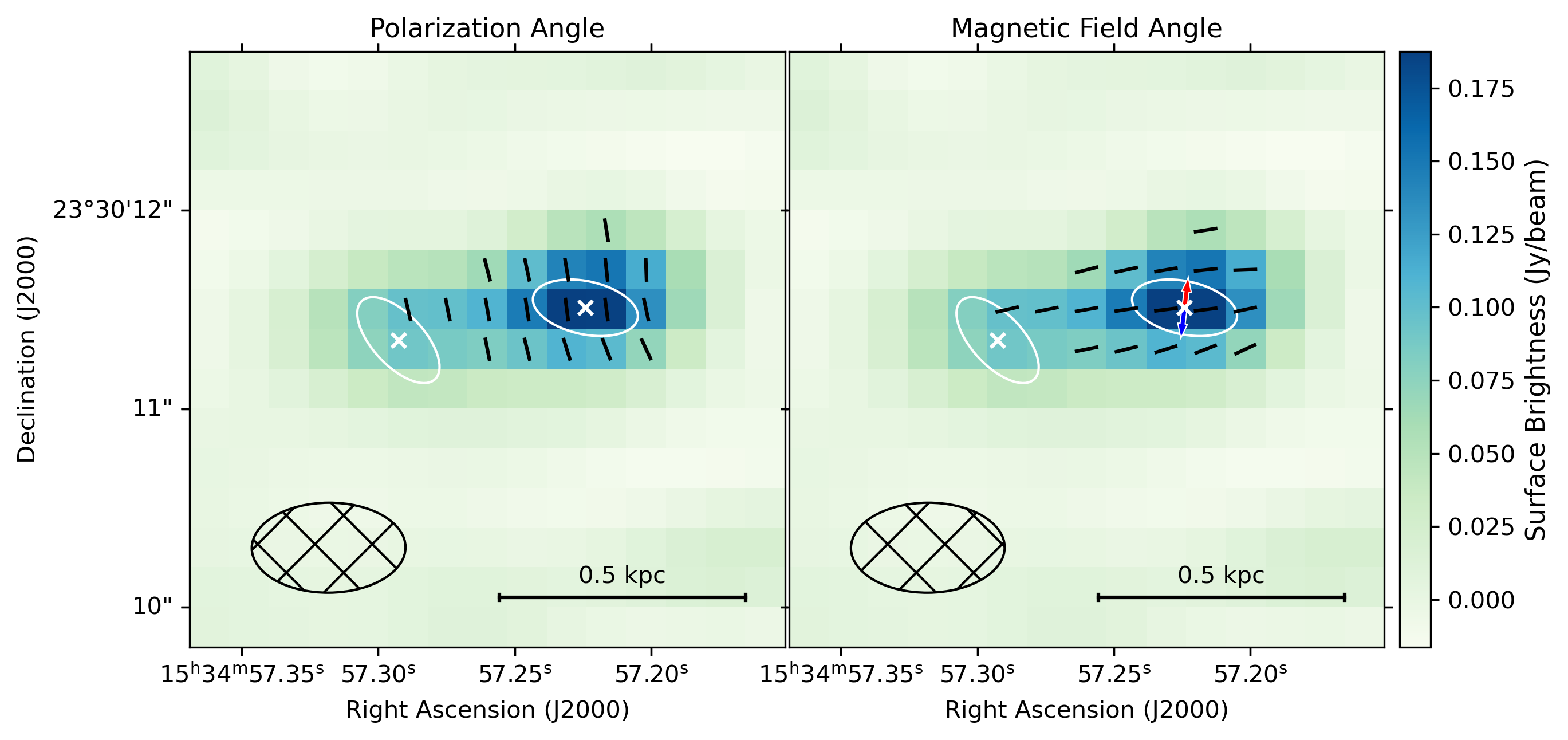}
\caption{Image showing the intensity of Arp 220 in the SMA continuum bands (colour) with polarization vectors overlaid (left). These are rotated by 90 degrees in the image to the right to show the direction of the magnetic field. Polarization is detected in the western nucleus at a peak significance of $6\,\sigma$. The beam is shown as a cross-hatched ellipse in the bottom left corner. The red ellipses represent the rotating molecular disks, while the positions of the nuclei are shown as red crosses. The blue-shifted and red-shifted lobes of the outflow from the western nucleus are marked; their length corresponds to the measured extent of the ouflow \citep{bm18}.}
\label{fig:arp220_pol}
\end{figure*}

\section{Results}

Our new subarcsecond resolution observations of Arp220 represent the most sensitive polarization study of this object to date. We show the continuum image resulting from our SMA observations overplotted with the polarization vectors derived from the Stokes Q and U images shown as lines, in Figure \ref{fig:arp220_pol}. We detect polarized dust emission in Arp220 for the first time, with a peak polarized flux intensity of $2.7\pm$0.45 mJy close to the position of the western nucleus. This represents a detection significance of $6\,\sigma$. This is the first detection of polarized dust emission from the nuclear regions of a ULIRG.

We make only a marginal detection of polarization associated with the eastern nucleus, with a polarized intensity of $1.2\pm$0.45 mJy, ie. a $2.6\, \sigma$ marginal detection at best. This is at least in part because the eastern nucleus is fainter in the continuum than the western nucleus by a factor of about 2. Emission from the two nuclei might thus be polarized at the same level, but we do not have the sensitivity to demonstrate this with the current observations. Our results are summarised in Table \ref{table:results}.

\begin{table*}
\begin{tabular}{cccc} \hline
&Peak Polarization&West Nucleus&East Nucleus\\ \hline
RA&15:34:57.25&15:34:57.22&15:34:57.29\\
DEC&23:30:11.70&23:30:11.52&23:30:11.33\\
Flux (mJy)&101$\pm$10&188$\pm$18&90$\pm$9\\
Polarized flux (mJy)&2.7$\pm$0.45&2.0$\pm$0.45&$<$2.55\\
Polarization fraction (per cent)&2.7&1.1&$<$2.8\\
Polarization angle&12.1$\pm$4.5&7.3$\pm$6.4\\
\hline
\end{tabular}
\caption{Basic parameters of the polarization results for the two nuclei and the peak of polarization. Where quoted, $3\,\sigma$ upper limits are given. Polarization angle refers to the electric-vector position angle. }
\label{table:results}
\end{table*}

\section{Discussion}

As well as the result of our observations and the detection of polarized dust emission associated with the western nucleus of Arp220, Figure \ref{fig:arp220_pol} also shows the inferred direction of the magnetic field derived from the dust polarization direction, since the magnetic field is expected to be orthogonal to the polarization direction. The figure also shows the locations of the two nuclei at the centre of Arp220 and the alignments of the dusty molecular gas disks around them.

%\color{red}

We do not resolve either the disc or the outflow in our observations, as shown in Figure~\ref{fig:arp220_pol}.  In non-interacting spiral galaxies, magnetic fields are typically seen to run along the spiral arms \citep[e.g.][]{j20}, consistent with the predictions of galactic dynamo models \citep{bn23}.  In an unresolved, non-interacting galactic disc, we might therefore expect to see an average magnetic field parallel to the major axis of the disc.  The magnetic field direction that we observe at peak polarization in the western nucleus of Arp220, $102.1^{\circ}\pm4.5^{\circ}$ lies between the position angle of the major axis of the western disc ($78^{\circ}$ E of N, \citealt{s17,b15}), and the angle subtended by the line between the eastern and western nuclei (124.7$^{\circ}$ E of N).  We thus hypothesise that a pre-interaction field driven by a galactic dynamo may be being distorted by the gravitational interaction between the two nuclei.  In the Antennae galaxies, for example, an ordered magnetic field connects the two interacting nuclei \citep{lr+23}, while elsewhere the magnetic field traces a relic spiral arm and a tidal tail.  The magnetic field in Arp220 may be evolving towards a similar state.

There is no indication that on the size scales that we observe, the magnetic field direction is correlated with the outflow direction.  This contrasts with observations of magnetic fields in more nearby galaxies with central outflows: in M82 \citep{j19,p21}, the magnetic field in the central starburst region is poloidal, and apparently entrained by the outflow.  Similar behaviour is seen in NGC 253 and NGC 2146 \citep{lr23}.  However, away from the central starburst regions of these galaxies, and in non-star-forming material in the centre of NGC 253, the magnetic field traces disc structure \citep{j19,lr23}.  If Arp220 were observed at higher resolution, e.g. using ALMA, we might expect to see the magnetic field being reordered by the outflow on smaller scales, as is seen in the central molecular zone of NGC 253, in which ALMA observations show that a magnetic field that is parallel to galactic structure in quiescent gas is reoriented to be parallel to the outflow direction in the star-forming superclusters on size scales of tens of pc in the centre of the galaxy \citep{lr23}.  

Arp220 is a system that is relatively early in its interaction; the western outflow is relatively compact ($\sim 100$\,pc), and dynamically young ($\sim 10^{5}$\,yr; \citealt{bm18}).  For comparison, the M82 `superwind' has an age of $\sim 3\times 10^{6}$\,yr, and extends 11\,kpc above and below the plane of the galaxy \citep{s98}, while the outflow from NGC 253 has an age $\geq 1$\,Myr \citep{w17} and is $\sim$ kpc-sized \citep[e.g.,][]{l23}, and the outflow from NGC 2146 has an age $\sim 10^{7}$\,yr and a height $\gtrsim 1$\,kpc above/below the plane of the galaxy \citep{t09}.  In order to identify the timescale on which major mergers, or the winds and outflows from starbursts triggered by these mergers, can cause significant disruption to a galactic dynamo field, polarization-sensitive observations of a range of more evolved systems, such as a larger sample of ULIRGs, are required.

\color{black}

Earlier consideration of the effects of polarized dust in ULIRGs and other far-IR luminous galaxies on B-mode CMB experiments, based on the SCUBA integrated polarization upper limits \citep{s07}, suggested that these sources would not present significant problems for {\em Planck}-like observations, based on the assumption that these sources would be polarized at a level of $\sim$ 1.5 per cent. Our detection of polarization in the western nucleus of Arp 220 at a level of 2.7$\pm$0.35 per cent, together with the increased sensitivity of a new generation of CMB instruments such as the Simons Observatory \cite{a19}, and other direct and statistical measurements of dust polarization in galaxies (eg. \citealt{b17}) suggests the need for a reconsideration of polarized dust emission in far-IR luminous galaxies as a polarization foreground, especially at higher frequencies.

\section{Conclusions}

We have observed the nuclear regions of the local ULIRG Arp220 with the SMA at a frequency of 340GHz and have detected polarized dust emission in western nucleus at a significance of $6\,\sigma$. This is the first time that polarized dust emission, thought to be the result of aspheric dust particles aligning with magnetic fields, has been detected in the nucleus of a ULIRG. We set upper limits to the polarized flux coming from the fainter, eastern nucleus. These limits, and a marginal $2.6\,\sigma$ polarized flux detection at the eastern nucleus' location, are consistent with a similar level of polarization to the western nucleus. Further, deeper observations with the SMA, or potentially ALMA, are needed before anything further can be said about polarization in the eastern nucleus. 

The magnetic field direction that we observe at peak polarization in the western nucleus of Arp220, $102.1^{\circ}\pm4.5^{\circ}$ E of N, lies between the position angle of the major axis of the western disc ($78^{\circ}$ E of N), and the angle subtended by the line between the eastern and western nuclei (125$^{\circ}$ E of N).  We thus hypothesise that a pre-interaction field driven by a galactic dynamo may be being distorted by the gravitational interaction between the two nuclei.  Unlike in more evolved starburst galaxies, there is no indication that, on the size scales that we observe, the magnetic field has been reordered by the central outflow, although such reordering might become apparent in higher-resolution observations.  In order to identify the timescale on which major mergers, or the winds and outflows from starbursts triggered by these mergers, can disrupt a galactic dynamo field, polarization-sensitive observations of a larger sample of ULIRGs are required.

While the observations described here deal with just a single target, the nearest and brightest ULRG, Arp220, they suggest that magnetic fields may play a significant roles in the processes underway in the innermost regions of major mergers. Observations in search of dust polarization in the inner regions of other local ULIRGs and other DSFGs are thus likely to bring new insights into these objects and how they evolve. ALMA is the ideal instrument for this since the next brightest local ULIRG to Arp220 is a factor of four or more fainter \citep{c18, c10}.
\\~\\
{\bf Acknowledgements}
The authors recognize and acknowledge the very significant role that Maunakea has always had for the indigenous Hawaiian community; we are fortunate to have been able to conduct our observations from its summit.
The Submillimeter Array is a joint project between the Smithsonian Astrophysical Observatory and the Academia Sinica Institute of Astronomy and Astrophysics and is funded by the Smithsonian Institution and the Academia Sinica. This work made use of Astropy\footnote{http://www.astropy.org} a community-developed core Python package and an ecosystem of tools and resources for astronomy \citep{astropy:2013, astropy:2022}. This work is funded in part by STFC, part of UKRI, through grant ST/S001468/1.  K.P. is a Royal Society University Research Fellow, supported by grant no. URF\textbackslash R1\textbackslash 211322.
\\~\\
{\bf Data Availability}
The data for this paper is available through the SMA archive at https://lweb.cfa.harvard.edu/sma-archive/.
\\~\\

\end{document}